\documentclass[letterpaper]{article}

\usepackage[T1]{fontenc}
\usepackage{tikz}
\usepackage{tabularx}
\usepackage{hyperref}
\usepackage{booktabs}
\usepackage{xr-hyper}
\usetikzlibrary{shapes.geometric, arrows.meta, positioning, calc}

\usepackage{geometry}
\usepackage{setspace}

\usepackage[style = chem-acs]{biblatex}
\usepackage{graphicx}
\usepackage{float}
\newfloat{scheme}{htbp}{los}
\floatname{scheme}{Scheme}
\floatname{chart}{Chart}
\newfloat{graph}{htbp}{loh}

\usepackage[version = 4]{mhchem} %

\usepackage{authblk}
\author[1,*]{Hrishi V. Harsora}
\author [1] {Ashutosh Krishna Amaram}
\author[1]{Raghavan Ranganathan}
\affil[1]{Department of Materials Engineering, Indian Institute of Technology Gandhinagar, Gandhinagar - 382355, India}
\author[2]{Tarun K. Agarwal}
\affil[2]{Department of Electrical Engineering, Indian Institute of Technology Gandhinagar, Gandhinagar - 382355, India}

\title{Atomistic Insights into MOCVD Growth of GaAs using a Closed Loop Autonomous System Combining Directed Reinforcement Learning and Kinetic Monte Carlo Simulations}
\date{*Email: hrishi.harsora@iitgn.ac.in}

\usepackage{pdfpages}
\begin{document}
\maketitle

\begin{abstract}
     \noindent Conventional process optimization of the Metal-Organic Chemical Vapor Deposition (MOCVD) growth parameters to achieve high-quality thin films relies on iterative, resource-intensive experimental runs. While predictive modeling can circumvent this cost and effort, simulating thin-film growth is constrained by the vastly different length and time scales of the deposition process. While continuum (e.g., Computational Fluid Dynamics (CFD)) and first-principles methods (e.g., Molecular Dynamics (MD)) struggle to balance experimental length/time scales and atomistic accuracy, Kinetic Monte Carlo (KMC) bridges this divide by resolving atomistic events whilst remaining tractable at experimental length and time scales. However, the accuracy of KMC relies on kinetic rate parameters that require costly density functional theory (DFT) calculations. In order to circumvent this, we present a framework that calibrates the rate parameters directly against experimental growth data. A forward KMC growth model is coupled in a closed loop with a Monte Carlo Tree Search (MCTS) optimizer that navigates the high-dimensional parameter space, with the search steered by the degree of rate control of each reaction rate parameter. The approach invokes no material-specific assumptions and hence, any deposition process whose forward kinetics can be reduced to a set of rate-controlled events is amenable to the same treatment. We demonstrate it for the MOCVD of GaAs, where the calibrated model captures the characteristic temperature dependence of the measured growth rate. The framework is also able to recover the three well known kinetic regimes characterized by change in the rate determining step from precursor chemisorption to desorption. As a result, the framework converts routinely measured growth data into a mechanistically interpretable model of the deposition process. Since the growth dynamics depend upon the parameters of the reaction rates and the experimental data, the framework is generalized and can be applied to a range of technologically important thin films.
\end{abstract}

\section*{Keywords}

Metal-Organic Chemical Vapor Deposition, Kinetic Monte Carlo, Bortz-Kalos-Lebowitz, Monte Carlo Tree Search

\newpage

\section{Introduction}

\noindent III-V semiconducting compounds, composed of group 13 and group 15 elements, are cornerstones of various important optical and electronic devices\supercite{MOKKA,SZENG,DELAL}. Their numerous applications stem from two crucial material properties - a direct band gap, which most of them possess, and the ability to form alloys with multiple components\supercite{CAPASSO,vurgaftman2001band}. The direct band gap allows efficient absorption and emission of light and, as a result, III-V compounds have formed a strong platform for the development of devices that interact with light\supercite{MOKKA,SZENG}. Light-Emitting Diodes (LEDs)\supercite{nakamura1994candela, schubert2006led}, multi-junction solar cells \supercite{king2007multijunction, geisz2020sixjunction, yamaguchi2021multijunction}, and solid-state lasers\supercite{alferov2001heterostructure, kroemer2001quasielectric} are some examples of such devices. Most of these compounds also exhibit high carrier mobility and excellent thermal conductivity, which has driven their application in high-power devices where high switching speed and quick thermal dissipation\supercite{DELAL, mishra2002algangan, mimura2002hemt} are required for sustaining performance. \\

\noindent Among the numerous III-V semiconducting compounds, Gallium Arsenide (GaAs) is a material which has been widely used for the development of the aforementioned devices. Due to its direct band gap ($E_g \approx 1.42$~eV at room temperature\supercite{vurgaftman2001band}) GaAs shows efficient absorption and emission of light\supercite{MOKKA,sturge1962optical} and, as a result, it is used to fabricate high-efficiency solar cells \supercite{metaferia2019gaas_hvpe,moon2016gaas_flexible,chen2019ultrathin_gaas} and devices like photo-detectors \supercite{kim2017gaas_photodetector,zumeit2022gaas_photodetector} and LEDs \supercite{yu2025gaas_led_xray,lin2019irled}. Several alloys, such as InGaP, are lattice-matched to GaAs and hence form sub-cells of efficient multi-junction solar cells, such as the InGaP/GaAs/Ge triple-junction cell\supercite{king2007multijunction,yamaguchi2021multijunction}. GaAs exhibits a high charge carrier mobility (about 8500~cm$^{2}$\,V$^{-1}$\,s$^{-1}$ for electrons at room temperature\supercite{adachi1985gaas}) and an excellent thermal conductivity which make it a strong candidate for high-power and high-frequency devices like HEMTs\supercite{HIYAMIZU1982455,tracy2019integrated_hemt}, Radio Frequency (RF) power amplifiers  \supercite{li2023kuband}, and Gunn diodes \supercite{gunn1963microwave}. GaAs substrates are also used as hosts for microwave integrated circuits, such as power amplifiers built from GaAs transistors\supercite{li2023kuband}. Moreover, GaAs shows negligible lattice mismatch with AlAs and other III-V compounds, which enables the realization of strain-free heterostructures with clean interfaces\supercite{marmalyuk2019heterostructures}. The lattice mismatch between GaAs and AlAs (their lattice constants are 5.653 and 5.661~\AA{} respectively) is less than 0.2\%\supercite{adachi1985gaas}. As a result, the Al$_x$Ga$_{1-x}$As alloy series can be grown on GaAs with negligible strain. In HEMTs, for example, the high-mobility channel is a two-dimensional electron gas confined at an abrupt AlGaAs/GaAs interface\supercite{HIYAMIZU1982455,mimura2002hemt}. Such structures are also employed in various devices like Vertical Cavity Surface Emitting Lasers (VCSELs)\supercite{HOUDRE1992267,234462}, whose mirrors are stacks of tens of alternating AlGaAs and GaAs layers of a precise thickness. All of these devices therefore require epitaxial layers whose thickness, composition and interfaces are controlled down to the atomic scale.  \\

\noindent The many applications of III-V compounds rely on thin films functioning either independently or as part of complex heterostructures which are composed of films with varying chemical compositions. These thin films are usually synthesized through vapor phase deposition techniques\supercite{venables2003nucleation_growth,stringfellow1998organometallic}, notable amongst them being Molecular Beam Epitaxy (MBE)\supercite{cho1975mbe,joyce1985mbe}, Close-Spaced Vapor Transport (CSVT)\supercite{greenaway2016csvt,perrier1988csvt}, Chemical Vapor Deposition (CVD)\supercite{metaferia2019gaas_hvpe,simon2019dhvpe}, Plasma-Enhanced CVD (PECVD)\supercite{snyders2023pecvd,singh2025fecvd} and MOCVD \supercite{stringfellow1998organometallic,Bradley2002,Heikman2002,Kaiander2006}. Although these processes differ widely in their approach towards deposition over a  substrate, they have a common methodology. Precursor compounds carrying the desired group 13 and 15 elements are introduced into a reaction chamber as vapors. The reactors are tightly regulated by precisely controlling parameters such as temperature, pressure and gas flow. The vapors then react in the gas phase forming deposition products which adsorb and condense on the substrate, thus resulting in the formation of a thin film\supercite{denbaars1986decomposition}. Among the numerous techniques, MOCVD is one of the most extensively utilized \supercite{manasevit1968gaas,stringfellow1998organometallic}. The process benefits from the organic nature of its precursors which, due to the versatility and isomerism inherent to organic ligands, result in many different reaction pathways to deposit a film with the same chemical composition\supercite{denbaars1986decomposition}. Furthermore, the metal-organic precursors require a lower thermal activation temperature to react within themselves and to adsorb onto the substrate, thus reducing the thermal budget of the process\supercite{denbaars1986decomposition,Makarenko2023}. This chemical flexibility and efficient precursor decomposition has led to widespread adoption of MOCVD for deposition of III-V films\supercite{manasevit1968gaas,Bradley2002,Heikman2002,Kaiander2006}. The deposition of GaAs has been carried out using many different precursors. However, Trimethylgallium (\ce{Ga(CH3)3}) and Arsine (\ce{AsH3}) are widely used due to their high volatility \supercite{Makarenko2023} thus eliminating the need for additional input of energy. \\

\noindent Thin films of GaAs and other III-V compounds are used either by themselves (in devices like solar cells and LEDs) or as heterostructures (in HEMTs and VCSELs). MOCVD is also the main route to other III-V materials: GaN and its alloys with In and Al, which underpin blue and white LEDs\supercite{nakamura1994candela,schubert2006led} and AlGaN/GaN HEMTs for high-power, high-frequency electronics\supercite{mishra2002algangan,Heikman2002}, and InGaAs quantum structures for long-wavelength lasers and VCSELs\supercite{Kaiander2006}. The performance of these devices is tied to the structural quality and chemical composition of the films \supercite{kuech1987movpe,stringfellow1998organometallic}. Defects such as dislocations, which form when a lattice-mismatched film relaxes beyond a critical thickness, produce trap states within the band gap that hamper absorption in optical devices \supercite{matthews1974misfit,shockley1952statistics,lang1974dlts}, while an atomically rough interface in heterostructures scatters charge carriers, reducing the efficiency of VCSELs and solar cells and the carrier mobility in HEMTs \supercite{sakaki1987interface,ando1982twodim}. Some devices also require films of a precise stoichiometry \supercite{PhysRevB.65.115203,Warren,electronics11020259}. Since these films are realized using MOCVD, their physical and chemical structure is tied to the process parameters used \supercite{kuech1987movpe,stringfellow1998organometallic}, and tuning these parameters demands repeated experimental runs that consume significant time and resources\supercite{he2022mocvd,wakabayashi2019ml}.\\

\noindent In order to avoid repeated experimental cycles, substantial effort has focused on computational modeling to optimize the process parameters\supercite{MIHOPOULOS1998733,he2022mocvd}.  At the reactor scale, computational fluid dynamics (CFD) has been used extensively to optimize the flow of precursors through the reactor and near the substrate\supercite{CHUANG2014254,MITROVIC2007323,BLACK1991241,5667717}, but it gives no insight into the microscopic sub-processes that build the film. These are resolved by Molecular Dynamics (MD)\supercite{yan2024atomistic,hong2018chemical,meng2012molecular,izumi2004development} and Density Functional Theory (DFT)\supercite{an2020ab,Timoshkin2001,Pickett}, but the very small time steps these methods require restrict them to small length and time scales\supercite{voter2002timescale}. Kinetic Monte Carlo (KMC) is a probabilistic simulation approach that sits between the macroscale and microscale methods\supercite{voter2007kmc}. In addition to providing atomistic insights\supercite{fphy,nano12173052,Gimenez2025}, KMC can also be scaled to large length and time scales. This is because it evolves the system through discrete transitions between stable configurations instead of resolving the electronic structure or atomic vibrations as is the case with DFT and MD respectively. Herein, we present a KMC model based on the Bortz-Kalos-Lebowitz (BKL) algorithm\supercite{BORTZ197510}, which models the homo-epitaxial growth of GaAs by considering the four fundamental microscopic processes of adsorption, desorption, surface decomposition and surface incorporation. Traditional KMC simulation approaches rely on high throughput DFT simulations for obtaining the energy barriers of the various microscopic sub-processes\supercite{andersen2019kmcguide,reuter2006fpkmc}. The task of setting up these simulations requires the investment of a significant amount of time and computational resources, in the form of costly DFT simulations. In order to circumvent this obstacle, the KMC modeling engine is coupled with a continuous Monte Carlo Tree Search (MCTS) optimizer\supercite{manna2022continuous,koneru2025silicene} in order to tune the parameters in the model. This coupled framework is demonstrated in the context of the MOCVD of GaAs in order to obtain an agreement between simulated and experimental results. 

\newpage

\section{Methodology}

\begin{figure}[!htbp]
    \centering
    \includegraphics[width=0.97\linewidth]{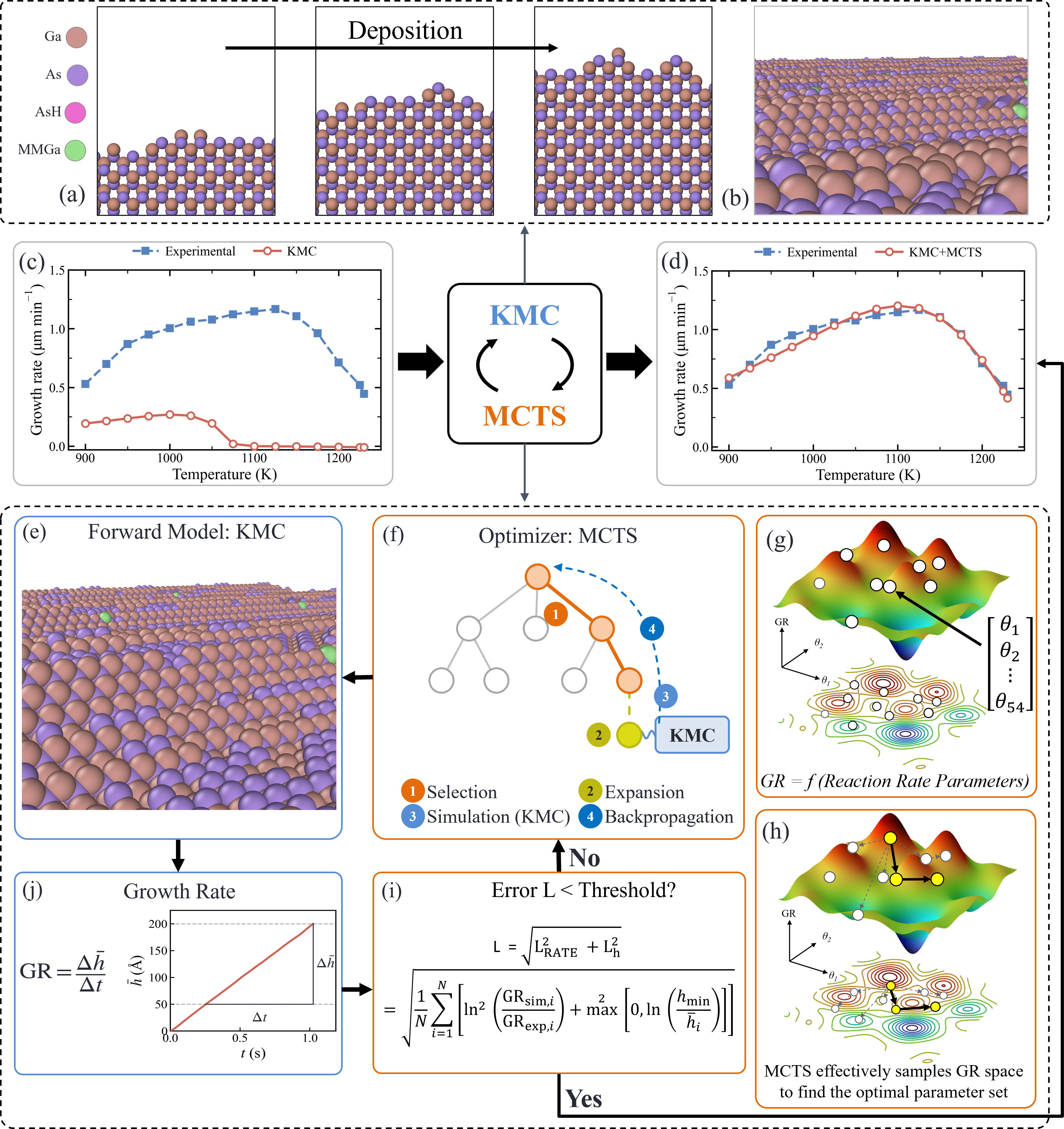}
    \caption{Closed-loop framework for calibrating the KMC model of GaAs MOCVD. (a)~Side view of the growing film at three successive times, 6, 14 and 22~\AA\ of mean thickness, on the Ga-terminated GaAs(001) substrate, with deposition proceeding from left to right. (b)~Surface of the grown film. (c,~d)~Growth rate versus temperature from experiment and simulation, (c)~with the default parameter set and (d)~with the calibrated set. (e)~The forward KMC model, which grows a film from a candidate parameter set. (f)~One MCTS iteration: selection, expansion, simulation with the KMC model, and back-propagation. (g,~h)~Schematic of the search for the optimal parameter set in a high dimensional space of parameters so as to achieve agreement between experimental and simulated growth rates. The surface is illustrative only. (g) shows the continuous high dimensional space where each point is a candidate parameter set, a vector of 54 parameters. (h) illustrates an MCTS tree that samples this space to search for the optimal parameter set. (i)~The error $L$ between the simulated and experimental growth rates, and the check that stops the search. (j)~The growth rate, read as the slope of the mean film height through the 50--200~\AA{} window.}
    \label{fig:Methodology}
\end{figure}

\noindent In order to model GaAs film growth, we adopt a framework consisting of two components - A forward model (figure-\ref{fig:Methodology}(e)) wherein growth is modeled with a Kinetic Monte Carlo scheme as depicted in figure-\ref{fig:Methodology}(a), (b) and an optimizer to find the kinetic parameters corresponding to experimental growth data (figure-\ref{fig:Methodology}(f)). A default set of kinetic parameters of the elementary surface events \supercite{MountziarisJensen1991} is used to simulate film growth using the forward model. Then the simulated growth rate (figure-\ref{fig:Methodology}(j)) is compared against experiment\supercite{Lin_GaAs_CVD} and the resultant mismatch (figure-\ref{fig:Methodology}(i)) is minimized by traversing the high dimensional space populated by numerous parameter sets using MCTS (illustrated in figure-\ref{fig:Methodology}(g)-(h)).
 
 \noindent The experimental growth was measured during the MOCVD of GaAs on (001)-oriented substrates using trimethylgallium (TMGa) and arsine (\ce{AsH3}) as precursors, at inlet partial pressures of $p_{\text{TMGa}}^{\circ} = 6.5 \times 10^{-4}$ atm and $p_{\text{AsH}_3}^{\circ} = 6 \times 10^{-3}$ atm respectively. These are the process parameters used in the forward KMC model. The growth rate values (supplementary information, section S1) reveal three distinct kinetic regimes: LT (900--1000~K), MT (1025--1150~K) and HT (1175--1230~K). These are illustrated in supplementary Figure S1.1. In the LT regime, the growth rate increases with temperature, indicating that the process is limited by the kinetics of the reactions occurring on the substrate surface\supercite{ReepGhandhi1983,Lin_GaAs_CVD,Heinecke1984}. In the MT regime, it varies only weakly with temperature, forming a plateau during which the supply of precursors to the film's growth front is the rate-limiting step\supercite{ReepGhandhi1983,Ludowise1985,BLACK1991241}. At high temperatures, the growth rate falls steeply as the deposited GaAs begins to decompose and desorb due to excessive thermal energy \supercite{ReepGhandhi1983,Ludowise1985}. To achieve agreement with experimental data, the two components, i.e., a forward Kinetic Monte Carlo (KMC) engine and a Monte Carlo Tree Search (MCTS) optimizer operate in a closed feedback loop as depicted in figure-\ref{fig:Methodology}. The KMC model uses a candidate parameter set to simulate film growth. As a result, the growth rate and behavior is a function of the parameters used. In order to achieve agreement with experimental data, these parameters are hence calibrated. Mathematically, this task reduces to a search for the optimal parameter set in a high dimensional space for which the MCTS optimizer is used. The specific mechanics of both components are detailed in the following subsections.
\subsection{Forward Model: Kinetic Monte Carlo (KMC)}

    \noindent Kinetic Monte Carlo (KMC) is a probabilistic method which is used for evolving systems through time. It utilizes random numbers drawn from a uniform distribution to select events from a predefined set to evolve the system and the simulation clock. It has been used extensively for modelling thin film deposition \supercite{DFTSLOP, DFTSLOP231, DFTSLOP3, DFTSLOP4, DFTSLOP5,nano12173052,Gimenez2025,BattailSrolovitzButler1997}. In our framework it serves as a forward model for film growth, as depicted in figure-\ref{fig:Methodology}(e).

\noindent Thin film deposition occurs through many microscopic sub-processes. Between these, the atoms of the system undergo rapid thermal vibrations about a local minimum of the energy that do not change the configuration \supercite{FichthornWeinberg1991,voter2007kmc}. These vibrations occur on timescales orders of magnitude shorter than the events that actually change the configuration. In the context of film deposition, these are events such as adsorption, surface decomposition and surface incorporation. Resolving the atomic vibrations explicitly is not only computationally expensive but also unnecessary if the goal is to simulate a system for a time orders of magnitude longer. The KMC method ignores such fast vibrations and advances the system directly from one minimum to the next whilst treating each barrier crossing as a discrete event \supercite{BORTZ197510}. This coarse-graining is what allows the simulation to reach experimentally relevant length and time scales\supercite{andersen2019kmcguide,fphy}.

\subsubsection{Description of the KMC model: on-lattice}

\noindent  In our framework, the KMC model used is on-lattice which means the chemical reactions occur at specific sites on a GaAs lattice. Each particle occupies a  crystallographic site on the zinc-blende lattice. This on-lattice description makes the set of chemical reactions (events) at any site and their rates a function of the local atomic environment. The simulation is carried out on a 10~nm $\times$ 10~nm cell. The cell size is one of several modelling choices that define the calibration problem: since the rate parameters are calibrated on this cell, a different choice would in general lead to a different, equally valid parameter set. The mechanistic interpretation drawn in the Results does not rest on individual parameter values but on which reactions control growth in each temperature regime, as discussed in the Conclusions. Periodic boundary conditions are applied in the two in-plane directions. As a result, this finite cell represents an effectively infinite growth surface while the film grows in the (001) out-of-plane direction.\\

\noindent The KMC forward model evolves the system using the Bortz-Kalos-Lebowitz (BKL) algorithm \supercite{BORTZ197510} (detailed in supplementary information, section S2). It requires prior description of all chemical events possible in the system along with their rates. In our framework, the chemical events considered are (1) adsorption of precursors, (2) desorption of surface species, (3) surface decomposition of adsorbate and (4) surface bond formation. The adsorption reaction is conveyed using a Hertz--Knudsen formulation\supercite{Hertz1882,Knudsen1915}. This links precursor flux to its partial pressure and the sticking coefficients of the precursors at the terraces and step edges of the film\supercite{stringfellow1998organometallic,BCF1951,venables2003nucleation_growth}. The events of desorption, surface decomposition, and surface incorporation are considered using an Arrhenius formulation\supercite{Arrhenius1889,voter2007kmc}. All energy barriers are given as $E/k_B$ in kelvin (supplementary information, section S3). The growth behavior that emerges during the simulation is set by the rates of these reactions. More concretely, it is a function of the reaction rate parameters. \\ 

\noindent In our framework, the system evolves using two random numbers drawn from a uniform distribution of the interval [0,1). This is in accordance with the formalism of the BKL algorithm\supercite{BORTZ197510,voter2007kmc}. At each iteration, the total rate $R_{\text{tot}} = \sum_{j=1}^{N} R_j$ is evaluated over all possible events on the surface, and the first random number (RN1) selects the event $m$ as the smallest index satisfying $\sum_{j=1}^{m} R_j \geq \mathrm{RN1}\,R_{\text{tot}}$, so that each event is chosen in proportion to its rate. In practice the engine makes this choice faster. All events of one reaction type share the same rate, so the random number first selects a reaction type, in proportion to the summed rate of its events, and then the event within it. Each event is still chosen with probability $R_j/R_{\text{tot}}$, but without summing over every event at every step (supplementary information, section S2). The second random number (RN2) advances the simulation clock by the residence time $\Delta t = -\ln(\mathrm{RN2})/R_{\text{tot}}$, so that configurations resulting in a higher total rate advance time by smaller increments. Then the system and simulation clock is updated to reflect the chemical reaction. This algorithm is illustrated with all mathematical equations in the supplementary information, section S2.

\subsubsection{Chemical reactions: events}

\begin{figure}[!ht]
    \centering
    \includegraphics[width=1\linewidth]{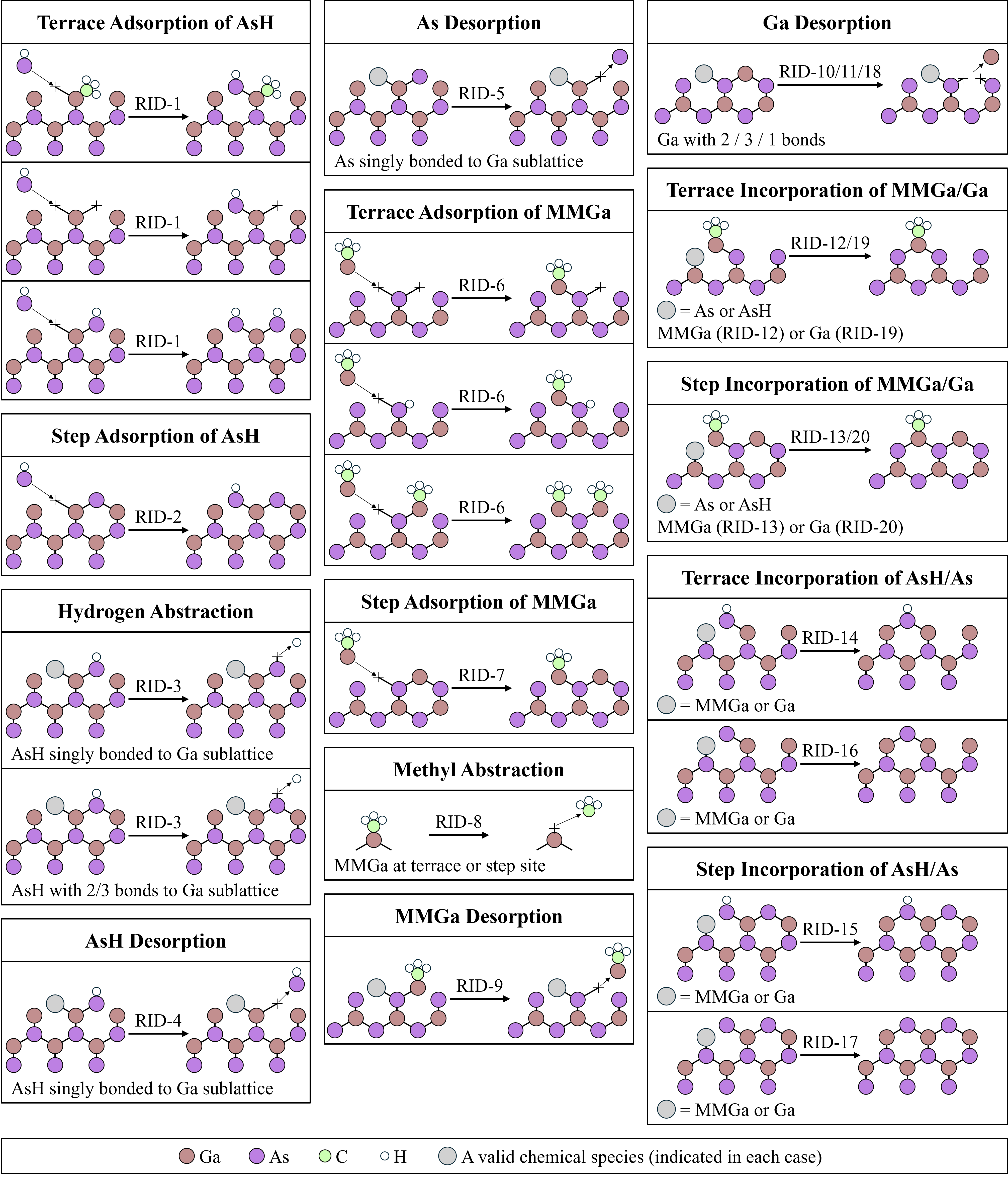}
    \caption{Illustrations of events considered by the KMC forward model. Adsorption attaches the decomposed precursors MMGa and AsH to the surface; surface decomposition strips the remaining ligands; desorption removes adsorbed species, their partially reacted intermediates and weakly bonded film atoms to the vapour; and surface incorporation forms Ga-As bonds at terrace and step sites.}
    \label{fig:Events}
\end{figure}

\noindent In this framework, the precursors considered for GaAs growth are trimethylgallium (\ce{Ga(CH3)3}) and arsine (\ce{AsH3}). However, these parent molecules do not reach the growth surface intact and have been known to undergo dissociation in the vapor phase itself \supercite{MountziarisJensen1991}. Accordingly, the forward model considers the decomposed products of monomethylgallium (MMGa, \ce{Ga(CH3)}) and arsenic monohydride (\ce{AsH}) as the effective growth precursors \supercite{Larsen1990} that actually adsorb on the surface. Twenty distinct surface reactions are considered: 8 are taken from Mountziaris et al. \supercite{MountziarisJensen1991} (reactions 3--5, 8, 9, 12, 14 and 16) and 12 are introduced in this work (reactions 1, 2, 6, 7, 10, 11, 13, 15, 17 and 18--20; Table~\ref{tab:energetics_provenance}). Thin film deposition then takes place by treating these reactions as discrete events and evolving the system over time by choosing one event after another, in accordance with the BKL formulation. The events are illustrated in figure~\ref{fig:Events}. These events fall into four elementary classes: adsorption, surface decomposition, desorption, and surface incorporation, as explained below. 

\vspace{0.2cm}

\noindent \textbf{(1) Adsorption} 

\vspace{0.05cm}

\noindent The adsorption event is mathematically described using a Hertz--Knudsen formulation\supercite{Hertz1882,Knudsen1915}. The growth front of a film consists of terrace and step regions and adsorption over both such sites can occur. In this model, a precursor may adsorb either onto a terrace (reactions 1 and 6 in figure~\ref{fig:Events}) or against a step edge (reactions 2 and 7). Adsorption on terrace and step sites is treated as two distinct events, since a precursor adsorbing at a step forms more bonds with the film and is therefore more stable.

\begin{equation}
k_{\mathrm{ads},i} = \left( \frac{P_i}{\sqrt{2\pi M_i k_B T}} \right) \cdot A_{\mathrm{site}} \cdot \left( S_0^{\mathrm{site}} \exp\left( -\frac{E_{a,\mathrm{chem}(i)}}{k_B T} \right) \right) ;\quad i = \mathrm{MMGa}/\mathrm{AsH} \text{ and } \mathrm{site} = \mathrm{terrace}/\mathrm{step}
\label{eq:adsorption}
\end{equation}

\noindent In Equation~\ref{eq:adsorption}, the first factor is the impingement flux of the precursor molecules \(i\) (\(i = \mathrm{MMGa}\) or \(\mathrm{AsH}\)) set by the partial pressure \(P_i\), molecular mass \(M_i\), and temperature \(T\). \(A_{\mathrm{site}}\) is the surface area associated with a single adsorption site. This converts flux to a rate at which molecules arrive at a site. Not every impinging molecule sticks to the surface. Hence, a sticking probability is associated with this impingement. Since attachment is thermally activated, this term takes an Arrhenius form, combining a baseline sticking coefficient \(S_0^{\mathrm{site}}\) (site = terrace or step) with an exponential factor governed by a chemisorption barrier \(E_{a,\mathrm{chem}(i)}\).

\vspace{0.2cm}

\noindent \textbf{(2) Surface Decomposition}

\vspace{0.05cm}

\noindent Following adsorption of a precursor at the appropriate site, reactions 3 and 8 illustrated in figure-\ref{fig:Events} account for their decomposition on the surface. These are mathematically captured using an Arrhenius formulation given below wherein $A_i$, $E_{a,i}$ and $\beta$ are the prefactor, energy barrier and temperature exponent respectively 

\begin{equation}
k_{\mathrm{dec},i} = A_i\, T^{\beta_i} \exp\!\left( -\frac{E_{a,i}}{k_B T} \right)
\label{eq:decomposition}
\end{equation}

\noindent Hydrogen is removed from the arsenic-bearing species (reaction 3 in figure~\ref{fig:Events}), while the methyl group is removed from gallium-bearing species (reaction 8). Each step exposes a dangling bond and thereby makes the species available for further bond formation. 

\vspace{0.2cm}

\noindent \textbf{(3) Desorption}

\vspace{0.05cm}

\noindent The desorption class of events accounts for the possible removal of adsorbates from the surface. This results in dangling bonds, thus making the surface more reactive. It acts not only on the incoming precursors, MMGa and AsH, but also on their decomposition products, Ga and As, so that a species may leave the surface at any stage (reactions 4, 5, 9, 10, 11 and 18 in figure~\ref{fig:Events}). The desorption rate is modelled using an Arrhenius formulation as shown below. 

\begin{equation}
k_{\mathrm{des},i} = A_i \exp\!\left( -\frac{E_{a,i}}{k_B T} \right)
\label{eq:desorption}
\end{equation}

\noindent $A_i$ and $E_{a,i}$ are the pre-exponential factor and energy barrier for desorption respectively. A well-coordinated surface atom has a higher energy barrier and hence a low tendency to desorb.

\vspace{0.2cm}

\noindent \textbf{(4) Surface Incorporation}

\vspace{0.05cm}

\noindent Surface incorporation is the event wherein a gallium and an arsenic species form a bond and the adsorbate becomes a part of the growing film. Incorporation may occur on a terrace (reactions 12, 14, 16 and 19 in figure~\ref{fig:Events}) or at a step edge (reactions 13, 15, 17 and 20). The latter must proceed over a substantially lower barrier than on a terrace, as the incorporating atom is coordinated by a greater number of neighbors. This effect is captured using the Arrhenius formulation shown below.

\begin{equation}
k_{\mathrm{inc},i} = A_i \exp\!\left( -\frac{E_{a,i}}{k_B T} \right)
\label{eq:incorporation}
\end{equation}

\noindent Here, $A_i$ and $E_{a,i}$ are the pre-exponential factor and the energy barrier for bond formation.

\noindent Every reaction in the model belongs to one of these four elementary classes — adsorption, desorption, surface decomposition, and incorporation. These together comprise 20 distinct reactions, as the rate of an event depends not only on its class but also on the local bonding environment and the identity of the species involved.

\subsubsection{Energetics of events}
\noindent The kinetic parameters (Equations~(\ref{eq:adsorption})-(\ref{eq:incorporation})) set the rates of the chemical events. These rates control the likelihood of an event being selected during the KMC simulation. Hence, the growth behavior is dependent on the reaction rate parameters. A majority of the parameter values are adopted from Mountziaris et al. \supercite{MountziarisJensen1991}. However, three groups of reactions have no counterpart in the reference\supercite{MountziarisJensen1991}: the desorption of film atoms held by two or three bonds (reactions 10 and 11 in figure \ref{fig:Events}), the adsorption of the growth precursors AsH and MMGa on terrace and step sites (reactions 1, 2, 6 and 7), and the reactions of an adsorbed Ga (reactions 18--20), which take the values of the corresponding MMGa reactions. The provenance of all parameters is summarized in Table~\ref{tab:energetics_provenance}. 

\begin{table}[!ht]
\centering
\renewcommand{\arraystretch}{1.3}
\begin{tabularx}{\textwidth}{@{}X l l@{}}
\toprule
Event class (reactions) & Rate parameters & Source \\
\midrule
Decomposition / H-abstraction (3, 8) & $A$, $E_{a}$, $\beta$ & Mountziaris \emph{et al.}\supercite{MountziarisJensen1991} \\
Desorption --- AsH, As, MMGa (4, 5, 9) & $A$, $E_{a}$ & Mountziaris \emph{et al.}\supercite{MountziarisJensen1991} \\
Terrace incorporation (12, 14, 16) & $A$, $E_{a}$ & Mountziaris \emph{et al.}\supercite{MountziarisJensen1991} \\
Step incorporation (13, 15, 17) & $A$, $E_{a}$ & This Work \\
Adsorption (1, 2, 6, 7) & $S_0^{\mathrm{terrace}} \le S_0^{\mathrm{step}}$, $E_{a,\mathrm{chem}}$ & This Work \\
Film-atom desorption (10, 11) & $A$, $E_{a}$ & This Work \\
Adsorbed Ga (18--20) & $A$, $E_{a}$ & This Work \\
\bottomrule
\end{tabularx}
\caption{Provenance of the kinetic parameters entering the event rate expressions. The parameters of the events absent from the reference are introduced in the present work and calibrated by the optimizer. The complete set of starting values is given in supplementary information, section S3.}
\label{tab:energetics_provenance}
\end{table}

\noindent  The modifications shown in Table \ref{tab:energetics_provenance} have been introduced into the standard rate equations in order to capture certain crucial effects that impact the growth behavior. The reactions involving desorption of film atoms held by two or three bonds (reactions 10 and 11 in figure \ref{fig:Events}) have been introduced in order to accurately capture the growth characteristics at high temperatures where excessive thermal energy leads to the decomposition of the deposited film\supercite{Ingle1996,ReepGhandhi1983}. As established earlier, the growth front of a film is rarely flat\supercite{BarabasiStanley1995,venables2003nucleation_growth} and is populated with numerous steps and kinks. These arrangements facilitate film growth by allowing the growth precursors and other surface species to form multiple bonds whilst being incorporated\supercite{BurtonCabreraFrank1951,venables2003nucleation_growth}. This effect is captured by giving step sites their own sticking coefficients and lower incorporation barriers than terrace sites. Many previous applications of KMC for thin film growth\supercite{DFTSLOP, DFTSLOP231, DFTSLOP3, DFTSLOP4, DFTSLOP5, fphy,nano12173052,Gimenez2025,BattailSrolovitzButler1997} do not consider all of these effects together. \\

\noindent The growth behavior modelled by the KMC forward model is dependent directly on the reaction rate parameters, some of which have been introduced in this work and hence whose values are not known from literature. The conventional route, adopted in earlier applications of KMC to thin film growth \supercite{DFTSLOP, DFTSLOP231, DFTSLOP3, DFTSLOP4, DFTSLOP5} was to use first principles calculations to obtain the values. Activation barriers were obtained from nudged elastic band calculations, with prefactors from transition state theory. Apart from the excessive computational cost, this approach is prohibitive for two reasons. Firstly, the barriers are functions of the local coordination environment rather than single numbers, so the count of required calculations multiplies with the number of distinct site configurations sampled during growth. Furthermore, the residual error of a converged, DFT-derived energy barrier is amplified exponentially by the Arrhenius form in the rate expressions. We have therefore re-framed the problem. Rather than computing the reaction rate parameters and predicting the growth rate, we treat the parameter set as an unknown variable. Then, we search for a set of values that reproduces the experimental growth rate across the temperature range. Hence, the problem boils down to a search in a high dimensional parameter space for an optimal set. For this, we utilize the Monte Carlo Tree Search (MCTS) algorithm whose application in our framework is detailed in the following sections. 

\subsection{Optimizer - Monte Carlo Tree Search (MCTS)}

\subsubsection{Introduction}

\noindent MCTS is a reinforcement-learning algorithm which has been applied previously to sequential decision problems\supercite{10.1007/11871842_29,10.1007/978-3-540-75538-8_7} and more recently, to the parametrization of interatomic potentials\supercite{10.1021/acs.jctc.4c00639,10.1021/acs.jctc.5c01885}. Its broad applicability stems from the efficiency with which it searches high dimensional parameter spaces. It does so by growing a tree whose nodes are candidate parameter sets. The search in the parameter space narrows down as the tree grows. Neighborhoods of parameter sets that have scored well are exploited, while regions that remain poorly sampled are favored for evaluation, using a reward metric for each node (supplementary figure S4.1). This combination of a principled trade-off between exploration and exploitation with a gradient-free tree structure is what makes the algorithm both efficient and broadly applicable.

\noindent In our framework, MCTS is applied to an inverse problem: the high-dimensional space is the space of kinetic parameters which determine reaction rates in the KMC forward model, and hence the modelled growth behavior. The optimal solution is the parameter set that best reproduces the experimental growth rate data\supercite{Lin_GaAs_CVD}. MCTS and KMC are coupled into a single closed-loop framework (figure~\ref{fig:Methodology}(e)-(j)) wherein the optimizer proposes a candidate parameter set (figure~\ref{fig:Methodology}(f)) and the forward model simulates film growth using it (figure~\ref{fig:Methodology}(e)). The resultant growth rate (figure~\ref{fig:Methodology}(j)) is then computed and its mismatch with experimental data (figure-\ref{fig:Methodology}(i)) is then used to guide the optimizer for selection of the next candidate parameter set (figure-\ref{fig:Methodology}(f)).

\subsubsection{Application of Algorithm}

\begin{figure}[!ht]
    \centering
    \includegraphics[width=0.85\linewidth]{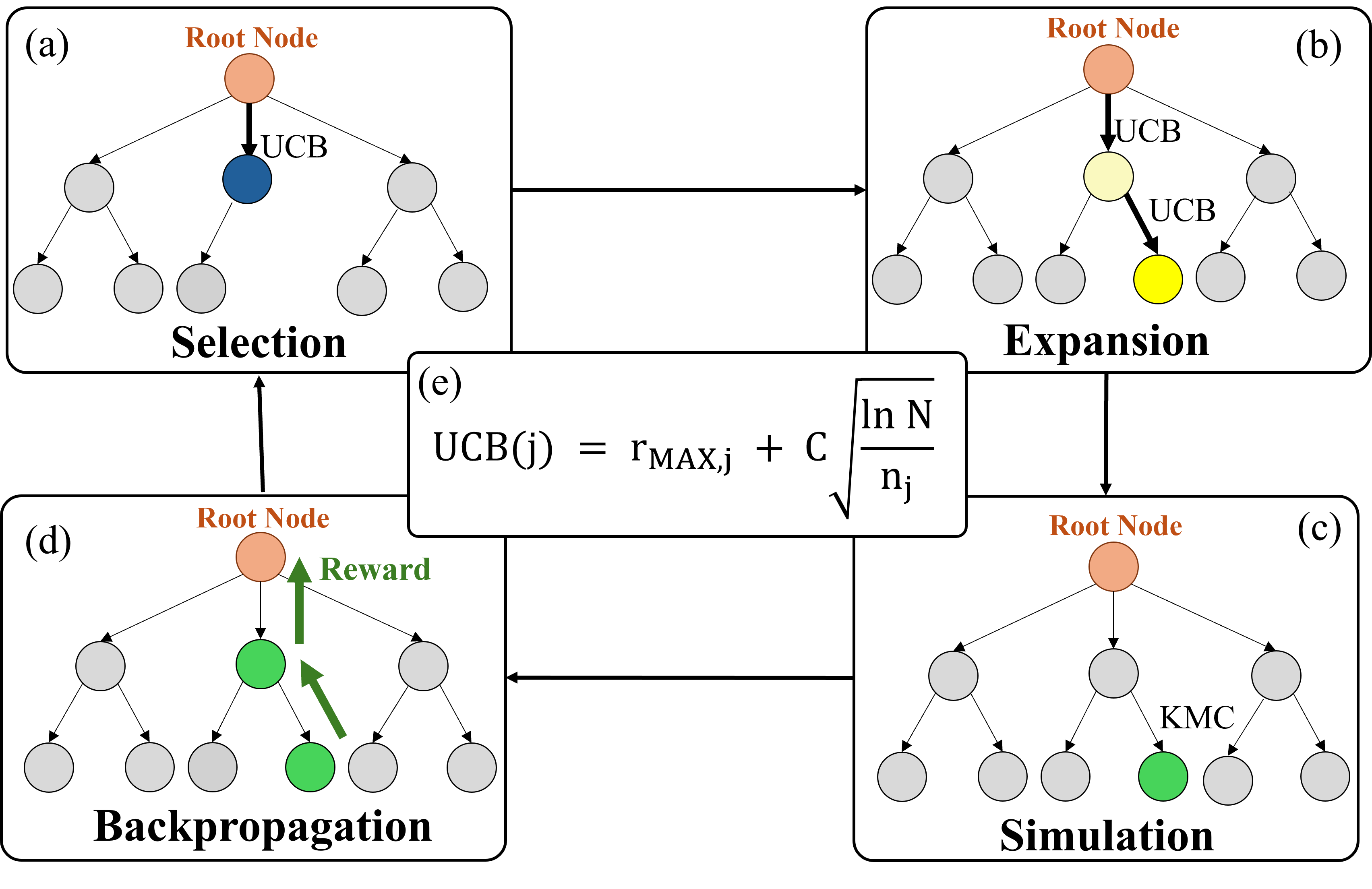}
    \caption{The four-stage MCTS optimization loop. Each node is a candidate parameter set and the root (orange) is the initial guess. In (a)~\emph{Selection}, the tree is descended from the root by repeatedly choosing the child with the highest upper-confidence-bound (UCB) score (panel (e)) until a leaf (dark blue) is reached. In (b)~\emph{Expansion}, a new child (yellow) is added by perturbing that node. In (c)~\emph{Simulation}, the new candidate is evaluated by the KMC forward model (green), returning a reward measuring agreement with experiment. In (d)~\emph{Backpropagation}, the reward is propagated back up to the root, updating every ancestor (green arrow).}
    \label{fig:MCTS_Workflow}
\end{figure}

\noindent Each node of the MCTS tree is a candidate parameter set, which sets the reaction rates in the forward model (Equations \ref{eq:adsorption} -- \ref{eq:incorporation}). The root is the initial guess, and each new node is a variation of its parent. The tree grows one node at a time through repeated cycles of four steps: selection, expansion, simulation and back-propagation (figure-\ref{fig:MCTS_Workflow}). Implementation details of the MCTS engine are given in supplementary information, section S4.

\noindent Each MCTS iteration faces a choice: Exploit a high rewarding region of the space by growing the existing branches and hence refining the solution or explore new regions of the space. MCTS balances between these choices using an upper-confidence bound (UCB) (figure-\ref{fig:MCTS_Workflow} (e)). Here, the first term represents exploitation:  $r_{\mathrm{max},j}$ is the best reward in the subtree rooted at node $j$, so nodes that match experiment are preferred. The second term represents exploration: $n_j$ is the number of visits to node $j$ and $N$ the number of evaluations so far. A rarely sampled branch is inflated, and the inflation falls as $n_j$ grows. The two choices are balanced using the exploration constant $C$. The value of $C$ is not assigned arbitrarily but evaluated at the start of the MCTS search from the error at the root and the degree of rate control of each parameter at the root. This ensures an appropriate value of $C$ and hence a balance between exploration and exploitation of the parameter search. This is detailed in supplementary section S4.   

\noindent Parameter sets are evaluated and scored by running the KMC forward model at the target temperature(s). The mismatch with respect to the experimental value is evaluated using the following error formulation:
\begin{equation}
L = \sqrt{\frac{1}{N_T}\sum_{T}\Big[L_{\mathrm{rate}}^{2}(T)+L_{h}^{2}(T)\Big]},
\label{eq:rmse}
\end{equation}
where the sum runs over the $N_T$ target temperature(s) and
\begin{equation}
L_{\mathrm{rate}}(T)=\ln\frac{\mathrm{GR}_{\mathrm{sim}}(T)}{\mathrm{GR}_{\mathrm{exp}}(T)},
\qquad
L_{h}(T)=\max\!\left(0,\ \ln\frac{h_{\min}}{h(T)}\right).
\label{eq:objective_terms}
\end{equation}

\noindent The error $L$ consists of the mismatch between the simulated and experimental GR at the target temperatures, $L_{\mathrm{rate}}(T)$, as well as a penalizing term $L_h(T)$. The latter is added to ensure that films of a reasonable thickness are deposited. Concretely, $L_h(T)$ is zero for films whose thickness reaches at least $h_{\min}=200$~\AA{} by the end of the simulation. A film that does not reach 200~\AA{} counts as a growth rate of zero, and a node whose films all fall short at any target temperature receives zero reward. The GR in $L_{\mathrm{rate}}(T)$ is evaluated from the time it takes for the thickness to increase from 50~\AA{} to 200~\AA. Moreover, the film height $h(T)$ in equation-\ref{eq:objective_terms} is the mean film height. For each film column of the lattice the topmost atom is taken, and those per-column heights are averaged over every column that carries one, measured from the starting substrate level. Thus, the formulation represents an average position of the growth surface, in agreement with what an in-situ ellipsometer (usually used for monitoring the film thickness during deposition) captures.

\noindent This error is then mapped onto a bounded reward which lies in the interval $(0,1]$: 
\begin{equation}
r = \exp(-L/\tau).
\label{eq:reward}
\end{equation}

\noindent This reward is stored on the new node. It is also back-propagated to the root and each node in the path updates its visit count $n_j$. If the new reward is larger it replaces $r_{\mathrm{max},j}$. The next
iteration starts at the root and scores siblings with those two numbers,
so a success (or a wasted visit) deep in the tree is visible when the
next child is chosen.

\noindent $L$ is measured in e-folds. Using a logarithm makes the error symmetric. Growing ten times too fast counts the same as ten times too slow and a film that does not grow is infinitely far from the target and scores 0 reward. A plain relative error
would rate it as merely fully wrong and the search would drift toward it. $\tau$ is a constant and is taken from the error at the root as $L_{\mathrm{root}}/3$. It scales the error so that nodes having similar error can still be differentiated.  

\noindent In a high dimensional parameter space like the one sampled in this work, there are many directions to perturb the parameter set of a parent node in order to spawn a child node and hence expand the MCTS tree. This problem is exacerbated when the space is continuous and small perturbations in the parameters lead to significant changes in the GR. In order to explore this parameter space efficiently and avoid spawning nodes randomly, the MCTS tree used in this work spawns child nodes following an informed decision made at its parent node. At the root, a degree of rate control (abbreviated as DRC and detailed in the next section) for each parameter in KMC is evaluated by introducing a small perturbation in one parameter at a time and observing its effect on the GR. A step planned from these slopes does not land exactly on the target, because the GR is not a linear function of the parameters. To account for this inherent curvature in the high dimensional space, the DRC of the parameters is updated from each measured move as the tree grows and thus the tree dynamically learns the direction towards the optimal parameter set. Further details about this are provided in supplementary section S4. 

\subsection{Optimization Strategy}

\noindent The MCTS search is not run over the full temperature range at once.
This is because the events that control the growth rate change with temperature, so
the DRC map is not the same in the LT, MT, and HT regimes. A single search over all three would move levers that correct one regime while degrading another. Such a search would need a larger MCTS tree. The optimization is therefore divided into four cumulative stages, one for each kinetic regime after a single-temperature start: 1125~K alone, where the measured growth rate is highest; then the LT, MT and HT regimes added in turn, with every stage keeping the temperatures fitted before it in its target set. Each stage searches the levers that control its new regime. At some stages, the root parameters produce no film: these are the first and last stages. The growth rate is then zero, and the DRC cannot be evaluated, since a small change in any one parameter does not restore growth. Each such stage therefore begins with a preliminary phase: the sensitivity of the film height to every parameter identifies those that promote growth, and these are adjusted together until the film grows (supplementary information, section S5). The parameters to be searched at each stage, the levers, are chosen from their degree of rate control (DRC): their effect on the growth rate per step, how far they can move it, and whether they move it in independent directions. For a parameter $\theta_k$ the DRC is evaluated using the following equation:

\begin{equation}
D_k
  \approx
  \frac{\ln\bigl(\mathrm{GR}_+/\mathrm{GR}_-\bigr)}
       {\delta_+-\delta_-},
\label{eq:drc_fd}
\end{equation}

\noindent where $\mathrm{GR}_{+}$ and $\mathrm{GR}_{-}$ are the growth rates obtained by increasing ($k\,e^{\delta_{+}}$) and decreasing ($k\,e^{\delta_{-}}$) the rate parameter respectively. The two steps are $\delta_{+}=+0.3$ and $\delta_{-}=-0.3$, in e-folds of
$\ln k$. $D_k$ is therefore $\mathrm{d}\ln\mathrm{GR}/\mathrm{d}\ln k$.
A positive value means that an increase in the rate raises the growth
rate and a negative value means that an increase in the rate lowers it.
A large $|D_k|$ identifies a lever for that regime while a value near zero
does not. The DRC is not measured for every parameter. It is measured only along the independent ones and a chain rule through $\ln k$ gives the sensitivity of all remaining parameters (supplementary information, section S4). 

\subsection{Mitigation of Noise}

\noindent KMC is a probabilistic technique, so independent simulations with the same parameter set give slightly different growth rates. Every growth rate used in the search is therefore the mean over an ensemble of independent simulations, large enough that its scatter is well below the differences the search acts on (supplementary information, section S5.1).

\section{Results and Discussion}

\subsection{Optimized Growth Behavior}

\begin{figure} [H]
    \centering
    \includegraphics[width=1\linewidth]{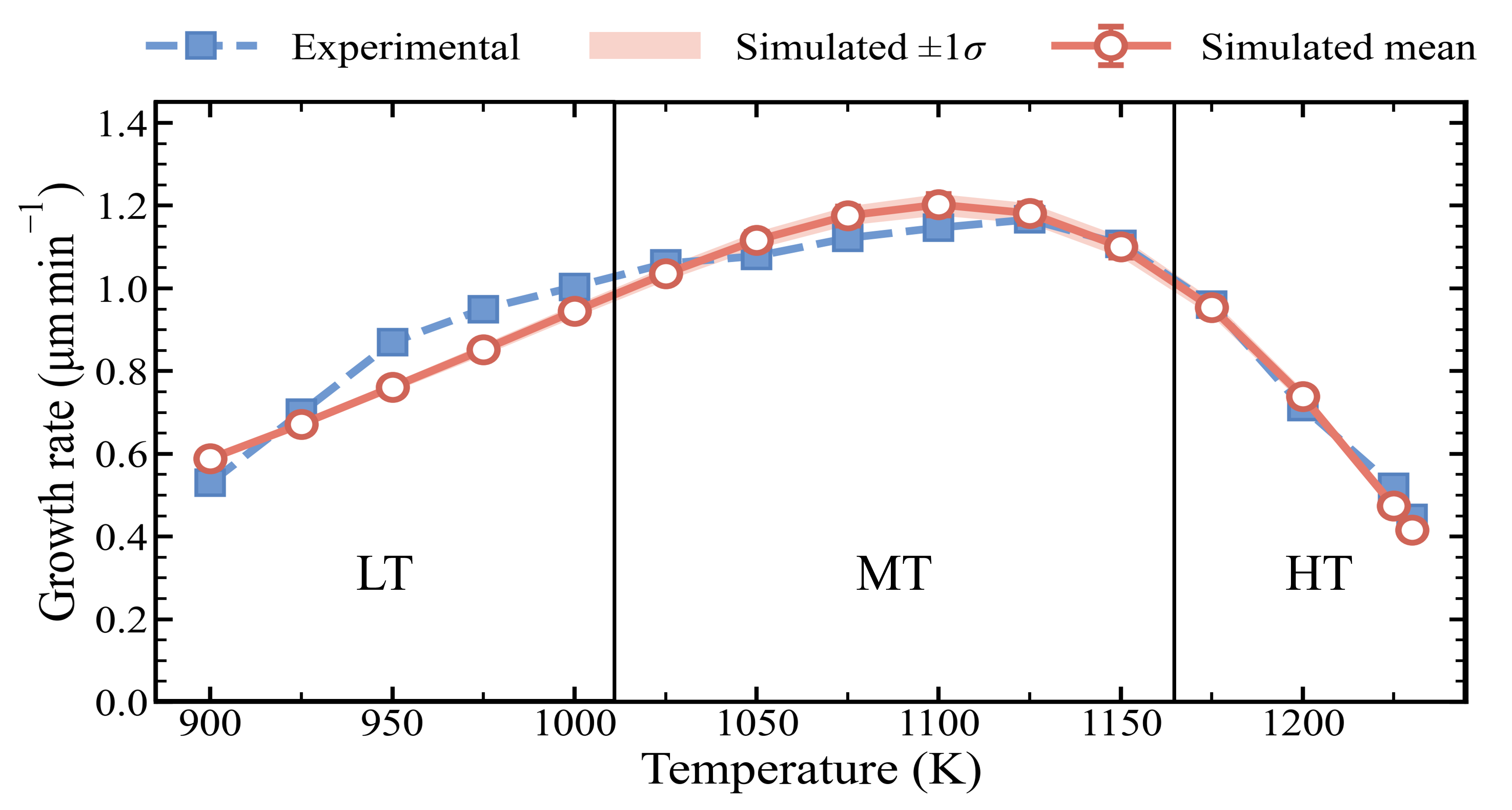}
    \caption{Optimized GaAs growth rate (GR) versus temperature (900--1230~K). Experimental data of Lin \emph{et al.} (blue squares) is compared with the mean over an ensemble of independent KMC simulations (red circles). The shaded band around the red curve denotes the inter-KMC noise discussed in previous sections. With the calibrated parameter set, the simulated GR reproduces the low-temperature rise, the peak near 1125~K, and the decrease at high-temperatures, as seen in the experimental data.}
    \label{fig:Calibrated}
\end{figure}

\noindent Figure~\ref{fig:Calibrated} compares the growth rate as simulated by the KMC model using the calibrated parameter set against the experimental growth rate reported by Lin \emph{et al.}\supercite{Lin_GaAs_CVD}. As a result of the optimization of the kinetic parameters, a single parameter set is now able to reproduce the essential features of the curve, whereas the default parameters resulted in underestimated growth (figure~\ref{fig:Methodology}(c)). The model recovers essential features of the three film deposition regimes: the rise through the low-temperature (LT) regime, the broad plateau in the mid temperature regime, and the sharp decrease in the high temperature (HT) regime. Although the behavior in each regime is driven by different sets of dominating reactions (see \emph{Kinetics from Growth Model}), a unified parameter set reconciles them all with a relative RMS deviation of the growth rate from experiment of \(\approx\) 7\%. The shaded band in figure~\ref{fig:Calibrated} is the film-to-film spread ($\pm 1\sigma$) of the growth rate among independent simulations run with the same parameter set, i.e., the inter-KMC noise introduced earlier in the Methodology section. It is not an error of the fit but a property of the kinetics, and it changes with temperature. This spread is about 1.2\% of the growth rate at 900~K and rises to about 2.2\% at and above the growth-rate peak. The noise is narrowest in the LT regime because the thermal energy is low. As a result, there are not many competing events and every independent KMC simulation leads to film deposition in much the same way as a result of which the simulations agree closely. The spread widens toward the peak and remains wide, relative to the growth rate, through the HT regime, where desorption of film atoms (reaction 10) becomes active. There, growth is the small remainder of two large and opposing flows, atoms arriving and atoms leaving, so small run-to-run differences in either flow are magnified in the net rate. The numerical values for the statistical spread are given in supplementary section S5.

\newpage
\subsection{Progressive Optimization }
\noindent The parameter set is optimized in four cumulative stages as depicted in figure~\ref{fig:Staged} below. Each stage adds one temperature regime to the set of target temperatures. At each stage, every target temperature, including those fitted in earlier stages, must lie within 10\% of experiment for the search to stop. The DRC (Degree of Rate Control as given in Equation~\ref{eq:drc_fd}) is measured at the beginning of each stage at each of the target temperatures. The parameters with the strongest and best-resolved control are then optimized by using MCTS. Across all 15 temperatures, the relative RMS deviation of the growth rate from experiment falls from 88\% for the default parameters to 51\%, 50\%, 39\% and finally 7\% at the end of the four stages.
\begin{figure}[!ht]
    \centering
    \includegraphics[width=1\linewidth]{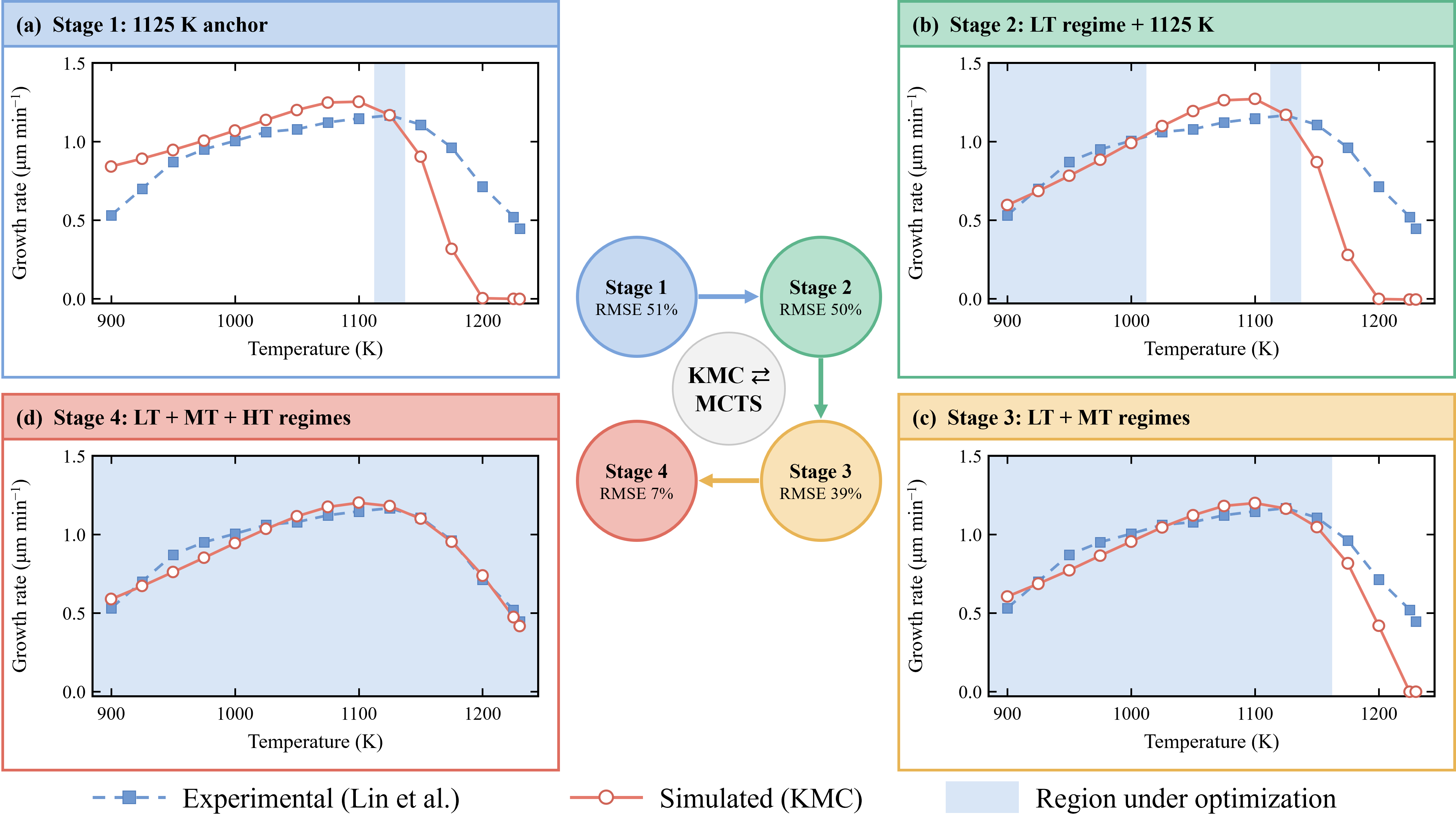}
    \caption{Stage-by-stage outcome of the optimization which proceeds through four cumulative stages: the 1125~K anchor, the LT regime + 1125~K, then further addition of the MT regime, and finally all three regimes (LT, MT, HT). Each inset shows the simulated growth rate (red) and the experimental curve of Lin \emph{et al.} (blue) at the end of the corresponding stage. The agreement tightens from one stage to the next, with the error (Equation~\ref{eq:rmse}) falling from $\approx 51\%$ at the anchor, where the tuned curve matches experiment only at 1125~K, to $\approx 7\%$ once all three regimes are fitted together.}
    \label{fig:Staged}
\end{figure}

\noindent The first stage considers 1125~K since the experimental growth rate here is the largest. With the default parameters no film grows at this temperature, so a growth rate cannot be measured and a DRC can hence not be established. A first phase therefore screens all the parameters for their effect on film height at 1125~K. It then calibrates the most effective parameters together until reasonable film thickness is seen. At the resulting parameter set, the DRC at 1125~K places almost all control on five quantities. These are the Ga precursor's sticking coefficients at the step and terrace sites (DRC 0.75 and 0.42), the As precursor's sticking coefficients (0.19 and 0.13), and the desorption of film atoms (reaction 10, $-0.50$). The other reactions have $|\mathrm{DRC}| < 0.012$, i.e., a very weak effect on the growth rate at this stage. MCTS tuned these five parameters and matched the experimental growth rate at 1125~K to within 0.3\%.

\noindent The low-temperature regime (900--1000~K) is added next resulting in a six temperature set. Growth here is also found to be controlled mainly by MMGa adsorption. The chemisorption barrier and sticking coefficients of this Ga precursor carry the largest DRC (about 0.95 for the chemisorption step at 900~K). Gallium desorption (reaction 18) has a small control at 900~K that vanishes at 1125~K ($-0.085$ against $-0.002$). It is retained because it acts on the low-temperature regime alone. The resulting set brings 925--1000~K and 1125~K within about 10\% of experiment, with 900~K 13\% high.

\noindent The addition of the mid-temperature regime (1025--1150~K) brings eleven temperatures into the target set. Toward the higher temperatures in this set, desorption of film atoms takes on a large share of the control, and the AsH chemisorption barrier also gains influence. With these parameters the plateau is reproduced within about 11\% from 925 to 1150~K, while the 900~K point remains 14\% high. To test whether this is a limitation of the search or of the model, the DRC was measured again near this solution and it was found that no parameter moved the growth rate at 900~K independently of 950~K. Every sticking coefficient was found to change the growth rates at both of these temperatures by nearly the same amount. 

\noindent The final stage adds the high-temperature regime (1175--1230~K) resulting in the full 15 temperature set. It was found that, with the parameter set adopted from stage 3, there was no appreciable thickness in the film at 1225 and 1230~K. This can be seen also in figure-\ref{fig:Staged}(c). Hence, as at 1125~K, a first phase here is executed to first get an appreciable film thickness. Investigation into the parameters for their control over the film height at 1225 - 1230~K identified the desorption of two-bonded film atoms as the process that leads to major decrease in the thickness. Hence, slowing its rate constant by a factor of $e^{0.45}$ ($\approx 1.6$) was found to be the smallest step that makes every temperature grow. Then, DRC measurement at this parameter set revealed that the desorption reactions are dominating the growth behavior in the high-temperature regime ($-1.0$ at 1200~K, $-2.6$ at 1230~K). Desorption of film atoms held by three bonds (reaction 11) acts only at 1225--1230~K ($-0.18$ and $-0.29$), which makes it the reaction which has the ability to shape the last two points without disturbing the temperature range of 1150--1200~K. It is added as a sixth parameter. MCTS on these six parameters gives the final set and a 7\% RMS deviation over all 15 temperatures is found. This set is the model shown in figure~\ref{fig:Calibrated} and analysed in figure~\ref{fig:Kinetics}.

\noindent The largest DRC values, in addition to the change in the value of each parameter, as done by MCTS, are detailed in supplementary information section S5.

\subsection{Kinetics from Growth Model }
\label{sec:kinetics}

\begin{figure} [p]
    \centering
    \includegraphics[width=0.86\linewidth]{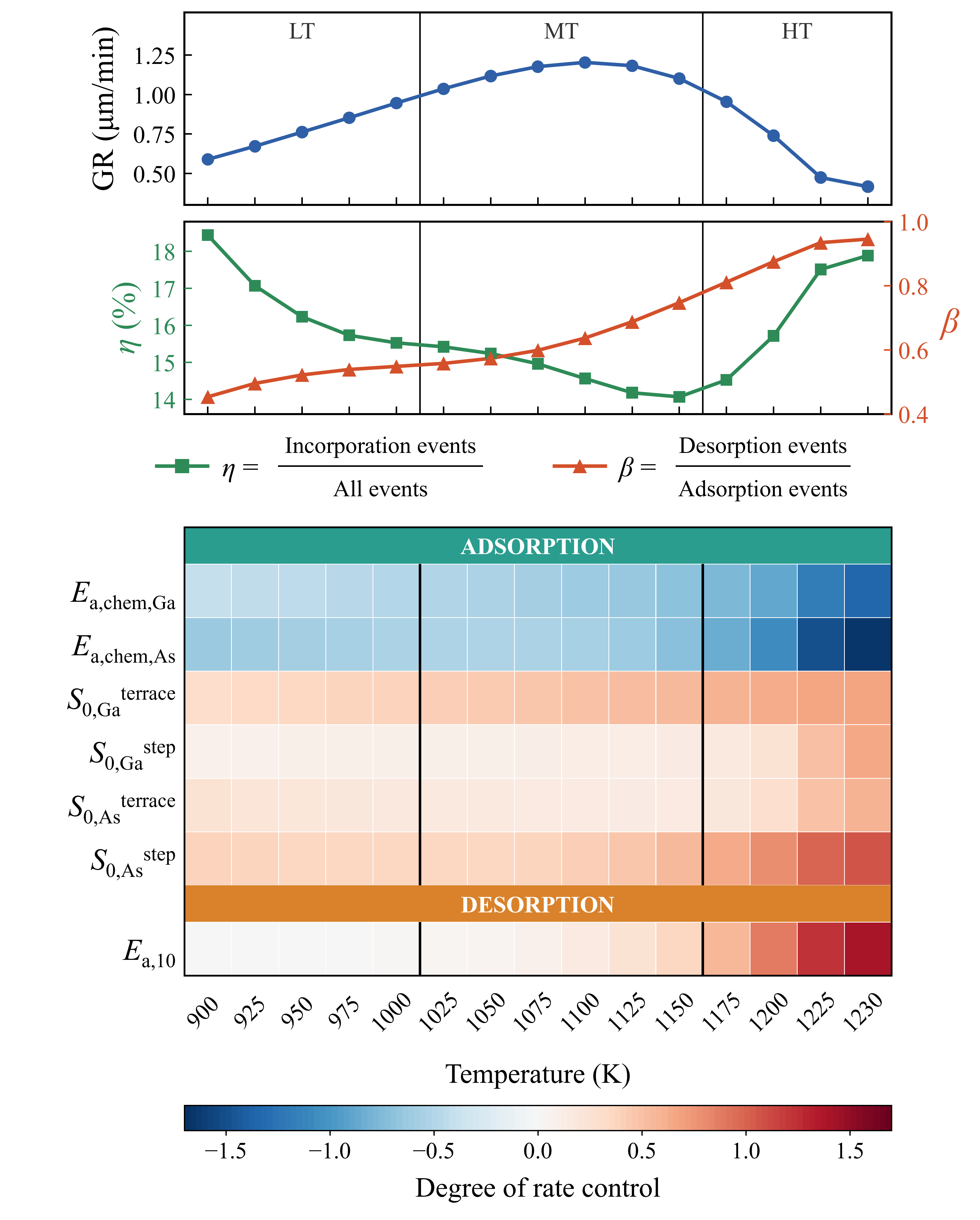}
    \caption{Temperature-resolved growth behavior of the optimized model; all panels share one temperature axis (900--1230~K), and the vertical dividers mark the LT/MT and MT/HT boundaries. (top)~Simulated growth rate. (middle)~Incorporation efficiency $\eta$ (left axis) and the desorption-to-adsorption ratio $\beta$ (right axis). (bottom)~Degree of rate control of every parameter that controls growth at some temperature ($|\mathrm{DRC}| \geq 0.1$; all others are below 0.1 everywhere), grouped into adsorption and desorption. For an activation energy the color is the response to raising that barrier; for a sticking coefficient, to raising that coefficient. Control rests on precursor adsorption throughout and is joined above 1150~K by desorption of two-bonded film atoms ($E_{\mathrm{a},10}$).}
    \label{fig:Kinetics}
\end{figure}

\noindent With the optimized parameters, the KMC model tracks the experimental growth rate across all three regimes. Hence, the reactions that produce such a behavior can now be investigated. Figure~\ref{fig:Kinetics} relates the rate-controlling processes, quantified by their DRC, to the growth rate and to two aggregate measures of surface activity.

\noindent The first measure is the incorporation efficiency $\eta$ which is the fraction of all executed events that are surface incorporations,
\begin{equation}
\eta = \frac{N_{\mathrm{inc}}}{N_{\mathrm{total}}},
\label{eq:eta}
\end{equation}
where $N_{\mathrm{inc}}$ is the number of incorporation events and $N_{\mathrm{total}}$ the total number of events. The second is the ratio of desorption to adsorption events,
\begin{equation}
\beta = \frac{f_{\mathrm{des}}}{f_{\mathrm{ads}}},
\label{eq:bratio}
\end{equation}
\noindent where $f_{\mathrm{des}}$ and $f_{\mathrm{ads}}$ are the fractions of executed events that are desorption and adsorption, respectively. $\beta$ measures how much of the material arriving at the surface leaves. 

\noindent In this section and in figure~\ref{fig:Kinetics}, the DRC of a reaction is quoted for its activation energy, that is, as the response of the growth rate to raising its barrier; it therefore has the opposite sign to the DRC of its rate constant (Equation~\ref{eq:drc_fd}) used in the previous section. Through the LT and MT regimes, growth is controlled by adsorption and the largest DRC belongs to the chemisorption barriers of the two precursors ($-0.41$ for MMGa and $-0.64$ for AsH at 900~K), followed by the sticking coefficients. No desorption reaction has appreciable control below 1050~K and it is found that the desorption of doubly bonded atoms (reaction 10) has a DRC of 0.02 at 1000~K. At low temperature the thermal energy is small compared with the chemisorption barriers, so only a fraction of arriving precursors actually chemisorb. Growth is therefore kinetically limited, and the rate climbs from 0.59~$\mu$m/min at 900~K to 0.95~$\mu$m/min at 1000~K. As the temperature rises, the barriers are overcome by a growing share of arrivals and the rise slows into the MT plateau, which peaks at 1.20~$\mu$m/min at 1100~K.

\noindent The passage from LT regime to MT regime is found to be gradual. The same adsorption parameters control both regimes but their DRC changes slowly over these temperatures.$\eta$ falls from 18.4\% to 14.1\% and $\beta$ rises from 0.45 to 0.75. At low temperature most desorption is not loss of film but it is adsorbed gallium leaving before it incorporates (reaction 18, about 10--12\% of all events), a fast exchange with no rate control ($|\mathrm{DRC}| < 0.02$).

\noindent At the boundary of the MT and HT regimes, the character of the surface changes. Desorption of doubly bonded film atoms, negligible at low temperature, grows from 0.5\% of events at 1025~K to 5.8\% at 1150~K and 15.5\% at 1230~K. Its DRC rises from 0.37 at 1150~K to 1.39 at 1230~K. Slowing this desorption raises the growth rate, which identifies decomposition of the deposited film as the origin of the experimental fall-off.

\noindent However, the adsorption parameters are still found to have control over the growth since their DRC grows as well with the increase in temperature. the DRC for the MMGa and AsH chemisorption barriers increases to $-1.35$ and $-1.65$ at 1230~K from $-0.69$ and $-0.71$ at 1150~K. At high temperature the net growth rate is the small difference between material arriving and material leaving, so it becomes sensitive to both of these processes.

\noindent The aggregate measures reflect the same picture. $\beta$ rises to 0.95 at 1230~K, so desorption balances adsorption. The number of events needed to grow a 200~\AA\ film rises more than tenfold from 900~K to 1230~K. $\eta$ passes through its minimum of 14.1\% at 1150~K and rises again to 17.9\% at 1230~K. At these temperatures many incorporation events get undone by the desorption that follows, so $\eta$ counts repeated incorporation of the same material, not productive growth.

\noindent The MT/HT boundary thus coincides with the onset of film-atom desorption as the minimum value of $\eta$ and the rise of the reaction~10 DRC both occur between 1150 and 1175~K. This boundary emerges from the optimized model rather than being imposed on it. The LT/MT boundary, by contrast, marks a gradual saturation of adsorption.

\subsection{Predicted Surface Morphology}

\begin{figure}[!ht]
    \centering
    \includegraphics[width=0.95\linewidth]{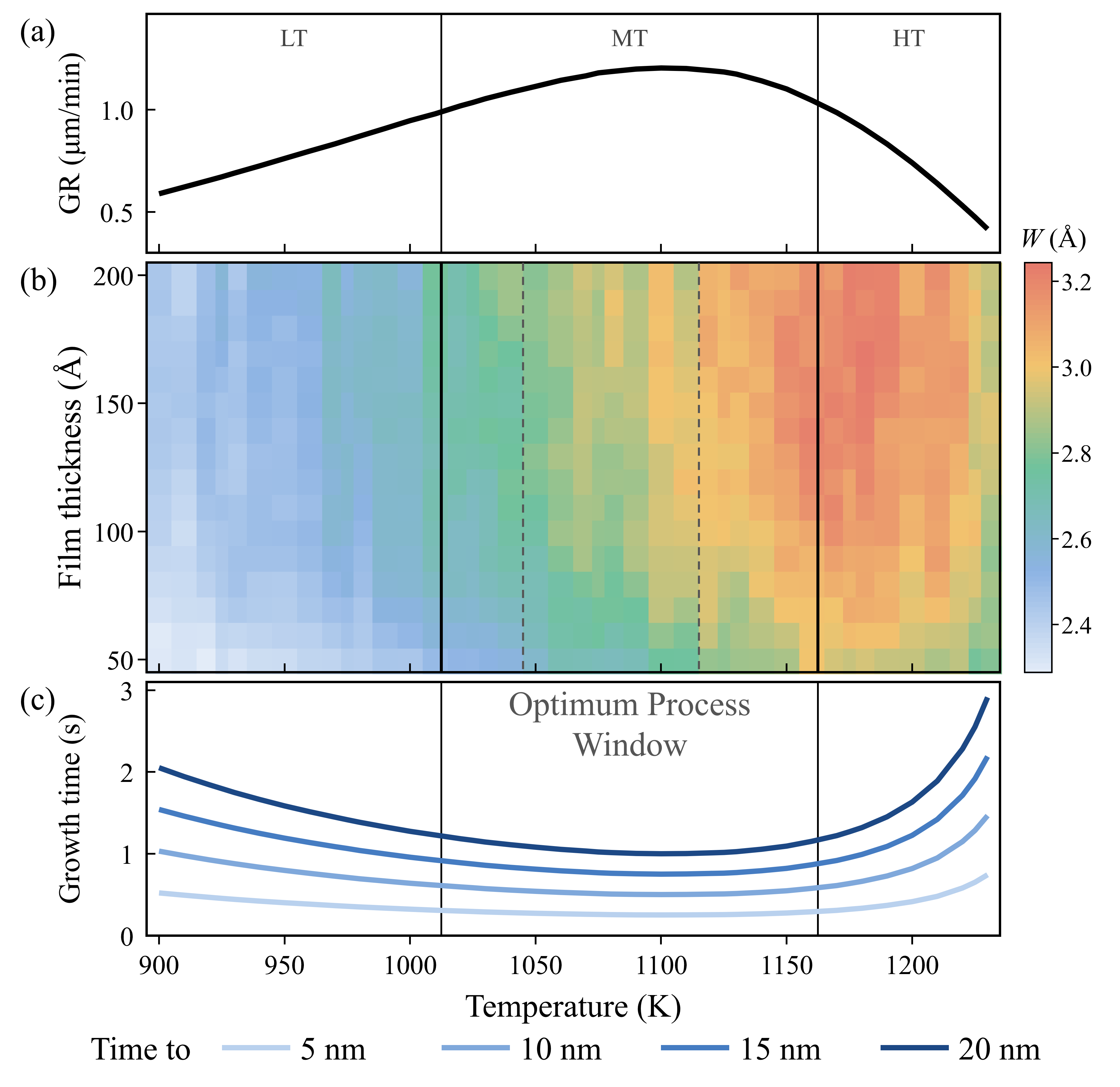}
    \caption{Surface morphology and deposition time at the final fit. (a)~Growth rate of the films against temperature. (b)~RMS roughness $W$ of the growth front over temperature (900 to 1230~K) and film thickness (50 to 200~\AA), where each cell is the mean over an ensemble of independent KMC simulations and $W$ is the standard deviation of the film height over the lattice columns. The maximum roughness ($\sim$1175~K) lies above the growth-rate peak (1100~K), at the onset of film-atom desorption (figure~\ref{fig:Kinetics}). (c)~Simulated time at which the films reach 50, 100, 150 and 200~\AA\ (5--20~nm) of mean thickness, against temperature. The three panels share the temperature axis. The dashed lines in (b) mark the process window (1050--1110~K), where growth is fastest; above 1110~K the films are rougher without growing faster.}
    \label{fig:Roughness}
\end{figure}

\noindent The parameters of the model were fitted to a single observable, the growth rate. The same simulations, however, resolve the growth front atom by atom, so the model also predicts other useful quantities, relevant to film deposition. The most relevant of these is the RMS roughness $W$ of the growth front which is here evaluated as the standard deviation of the film height over the lattice columns. Figure~\ref{fig:Roughness} maps $W$ at the final fit over the full temperature range (900--1230~K, 41 temperatures) and over film thickness (50--200~\AA, read every 10~\AA\ as each film grows), and because every simulation carries its own clock the figure also records, for the same films, the growth rate and the time needed to reach a given thickness. Each value is the mean over an ensemble of independent KMC simulations, as discussed before, to mitigate the inter-KMC noise. The atomistic snapshots behind the map are provided as XYZ trajectories in the project repository (Data and code availability).

\noindent Two trends are found to emerge. Firstly, the growth front at every temperature is found to be smooth as $W$ lies between 2.3 and 3.2~\AA (two to three atomic planes) and it saturates with thickness. It rises over the first $\sim$100~\AA\ of growth and is essentially constant thereafter, rather than continuing to grow. Secondly, the roughness is not a monotonic function of temperature, and its maximum does not coincide with the growth-rate peak. At 200~\AA, $W$ rises from 2.45~\AA\ at 900~K to a maximum of 3.23~\AA\ near 1175~K which is above the growth-rate peak at 1100~K, and then decreases to 2.92~\AA\ at 1230~K. The time to grow the film (figure~\ref{fig:Roughness}c) follows the growth rate, so that 200~\AA\ of film is deposited in 1.0~s at 1100~K but takes 2.0~s at 900~K and 2.9~s at 1230~K, and since the curves for 50, 100, 150 and 200~\AA\ run nearly parallel to one another, the time spent per unit of thickness at a given temperature is the same early and late in the deposition.

\noindent Interestingly, the temperature at which maximum roughness is found follows directly from the kinetics of figure~\ref{fig:Kinetics}. Through the LT and MT regimes, growth is adsorption-controlled and desorption of film atoms is negligible. The extent of roughening grows slowly with temperature. The maximum roughness falls between 1150~K and 1190~K, which is exactly where film-atom desorption (reaction 10) becomes a dominant reaction. Here, material is repeatedly incorporated and removed and the growth front turns over most actively while the surface is at its roughest. Above $\sim$1190~K the balance shifts further toward desorption ($\beta \to 0.95$), the net growth rate collapses, and the repeated removal of weakly bonded atoms smooths the front again. The maximum roughness is therefore a signature of the same desorption onset that produces the high-temperature fall-off as seen in the growth rate.

\noindent Taken together, panels (b) and (c) of figure~\ref{fig:Roughness} turn the fitted model into a process map, because they state that how long a film of a chosen thickness takes to grow and how rough it will be when the process is completed. The time needed to grow a 200~\AA\ film reaches its minimum, 1.00~s, at 1100~K, which is approximately the same temperature as the growth-rate peak. This growth time stays within 10\% of its minimum value from 1050 to 1150~K, but above 1110~K the films do not grow faster and also have a higher roughness, with $W$ of 3.06--3.09~\AA\ against 2.88~\AA\ at 1090--1110~K. The optimum process window is therefore 1050--1110~K (marked by dashed lines in figure~\ref{fig:Roughness}), where films grow in the shortest time. Further from it, growth slows quickly on both sides: a 200~\AA\ film takes 1.3--2.0~s between 900 and 1000~K and 2.9~s at 1230~K. The fastest deposition, at 1100~K, does not give the smoothest film, since a 200~\AA\ film grown there in 1.0~s has a roughness of about 2.9~\AA, whereas the same film grown between 950 and 1000~K takes 1.3 to 1.6~s and is smoother, with $W$ of about 2.5 to 2.6~\AA. The window therefore trades about 0.2--0.4~\AA\ of roughness for the shortest deposition time. Temperatures from 1110 to 1200~K are no faster than 1100~K and give the roughest films, and above 1200~K the deposition time rises steeply without any gain in smoothness, so that region offers nothing to a process. Where on this trade-off between throughput and surface quality a process should sit depends on the application, but the trade-off itself is now available, from a parameter set that was fitted to the growth rate alone.

\section{Conclusions}
\noindent In this work, we developed a self-optimizing Kinetic Monte Carlo (KMC) framework in which a forward atomistic growth model is coupled to a Monte Carlo Tree Search optimizer, and we demonstrated it for the MOCVD growth of GaAs. Rather than deriving the numerous, poorly constrained kinetic parameters from computationally expensive first-principles calculations,
the optimizer searches the high-dimensional kinetic parameter space, guided by the degree of rate control of each reaction and staged by temperature regime, to identify the parameter set that brings the simulated growth rate into agreement with experiment. The accuracy of the resulting model is therefore set not by the cost or fidelity of an underlying electronic-structure calculation but by the experimental data the optimizer is calibrated against, while the KMC description retains access to the length and time scales relevant to real film growth. Applied to GaAs, the calibrated model not only reproduces the measured growth-rate behavior but also recovers, from the degree-of-rate-control analysis, the three temperature regimes and the handover of control from precursor chemisorption to desorption of incorporated film atoms that underlies them, converting routinely measured growth data into a mechanistic picture of the process. The framework makes no assumptions specific to GaAs: any growth process whose forward kinetics can be expressed as a set of rate-controlled events, and for which experimental growth data are available, can be treated in the same way. It thus offers experimentalists a practical and broadly applicable route to optimizing deposition processes. Most importantly, this method replaces exhaustive trial-and-error growth runs and dedicated first-principles studies with a predictive, mechanistically interpretable model learned directly from the measurements they already collect.\\

\noindent The parameter set reported here should be read with the scope of the work in mind. The aim is not to settle the surface chemistry of GaAs but to demonstrate the utility of the closed-loop framework. The reaction set is deliberately minimal. A rich process like film growth admits far more atomic-scale events than any model can usefully carry, so only those events certain to occur are included. This set can certainly be expanded or reduced to incorporate necessary and/or missing chemistry. The parameter set that brings such a model into agreement with a measured growth rate is therefore conditional on four choices: which reactions and parameters are included, what experimental data is targeted for optimization, the rules each reaction follows (what counts as a terrace or a step, for instance), and also the size of the simulated system. As these choices change, the optimal set changes with them. This paper presents one such set together with the procedure that finds it. It must be noted that even for a fixed set of choices, the solution need not be unique because parameters can compensate for one another, so that different parameter sets reproduce the same growth-rate curve. In the third stage of the optimization, for example, a lower MMGa step sticking coefficient was largely offset by a lower MMGa chemisorption barrier. The individual parameter values should therefore not be read as measured quantities. What the analysis does establish is which reactions control the growth rate in each temperature regime, and it is this picture on which the mechanistic interpretation holds. Other researchers may reasonably make different choices, such as a different experimental dataset, a larger simulation cell or a richer reaction set. The framework applies to any such choice without modification and returns the parameter set appropriate to it.\\

\noindent The intended use case of the framework is a new film chemistry that is expensive to explore in ALD, CVD or PVD reactors. The essential reactions of such a system can be first identified through in-situ observations such as scanning electron microscopy (SEM), X-ray photon correlation spectroscopy (XPCS), etc. Only these reactions can then be built into the model with a set of rules and rate parameters that need be no better than educated guesses. The closed loop then finds the parameter set that agrees with in-situ ellipsometry or quartz-crystal-microbalance data (growth rate and flux) and with characterizations such as atomic force microscopy (AFM) surface scans for roughness, X-ray Reflectivity (XRR) data for density and X-ray Diffraction (XRD) for crystallinity. As a result, the arduous and expensive phase of first-principles rate calculations is thus bypassed completely.\\

\noindent Moreover, This closed loop framework can also help pave the way for agentic deposition systems. In the usual approach, a reactor runs long series of experimental runs to generate training data: films and their measured properties, such as growth rate (from ellipsometry), density (XRR), roughness (AFM) and crystallinity (XRD). This data is then used to train a surrogate model which learns to map process conditions to film properties. However, this map between process parameters and film properties is high dimensional because every film property depends on many interacting process parameters (temperature, pressure, precursor partial pressures and V/III ratio, flow rates, substrate orientation). Moreover, there can also be a day-to-day variability in the reactor conditions which may lead to the exact same process conditions not leading to films of the exact same characteristics. Thus, populating this space densely so that the surrogate model is able to learn the space to the fullest extent is difficult. A calibrated KMC model offers a different route. The model explores the process space in silico and predicts observables such as growth rate, roughness and density (using suitable digital twins of ellipsometer, AFM, and XRR). The reactor then grows only the films needed to test these predictions, and the measurements feed back to the MCTS optimizer, which recalibrates the model against them. Since KMC reaches experimentally relevant length and time scales it offers the unique advantage of being comparable with these measurements directly. The task of the reactor thus shifts from training a data-driven surrogate which requires massive amounts of data and correspondingly many experimental runs, to calibrating a digital twin of the process. Because the twin is built on the chemistry and physics of the surface reactions rather than learned from data alone, it is reasonable to assume that far fewer experimental runs would be needed for its calibration. This makes the framework a natural engine for agentic deposition systems.

\section{Data and code availability}
\noindent The code and data supporting this study will be made publicly available upon acceptance of this manuscript in a peer-reviewed journal.

\section{Acknowledgment}
\noindent The authors acknowledge the use of the PARAM Ananta Supercomputing facility at the Indian Institute of Technology Gandhinagar, established under the National Supercomputing Mission of the Government of India and supported by C-DAC, for the computational resources used in this work.

\printbibliography

\clearpage
\includepdf[pages=-]{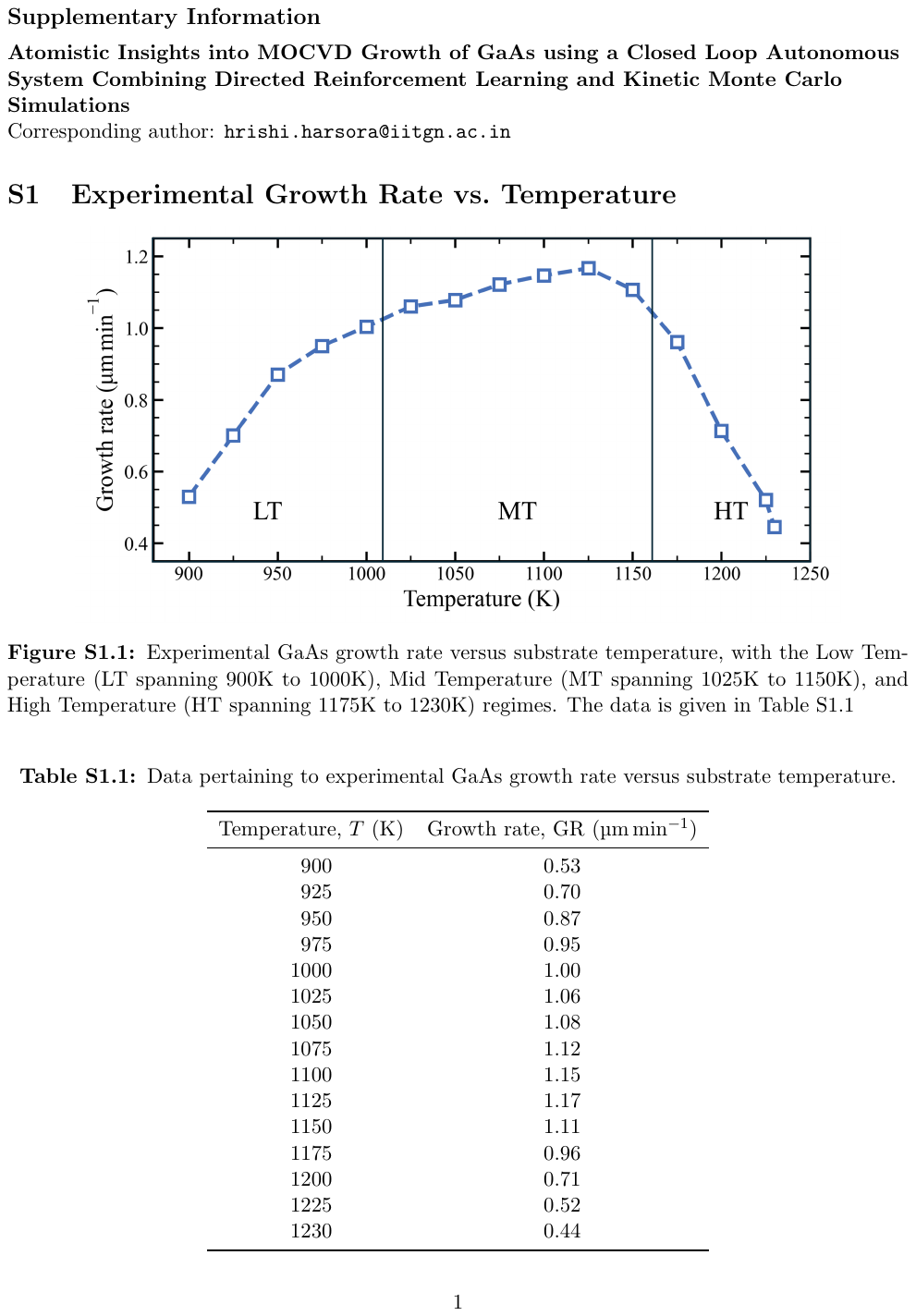}
\end{document}